\documentclass[aps,twocolumn,showpacs,floats,prl,10pt]{revtex4-1}

\usepackage{graphicx}
\usepackage{dcolumn}
\usepackage{bm}
\usepackage{color}
\usepackage{subfigure}

\usepackage[normalem]{ulem}

\begin{document}

\author{
  G. Ferr\'e, G. E. Astrakharchik, and J. Boronat
}

\affiliation{
Departament de F\'{i}sica i Enginyeria Nuclear, Universitat Polit\`{e}cnica de Catalunya,
Campus Nord B4-B5, E-08034, Barcelona, Spain
}

\title{Phase diagram of a quantum Coulomb wire}

\begin{abstract}
We report the quantum phase diagram of a one-dimensional Coulomb wire
obtained using the path integral Monte Carlo (PIMC) method. The exact
knowledge of the nodal points of this system  permits us to find the energy
in an exact way, solving the sign problem which spoils fermionic
calculations in higher dimensions. The results obtained allow for the
determination of the stability domain, in terms of density and
temperature,  of the one-dimensional Wigner crystal. At low temperatures,
the quantum wire reaches the quantum-degenerate regime, which is also
described by the diffusion Monte Carlo method. Increasing the temperature
the system transforms to  a classical Boltzmann gas which we simulate
using classical Monte Carlo. At large enough density, we identify a
one-dimensional ideal Fermi gas which remains quantum up to higher
temperatures than in two- and three-dimensional electron gases. The
obtained phase diagram as well as the  energetic and structural properties
of this system are relevant to experiments with electrons in  quantum wires
and to Coulomb ions in  one-dimensional confinement.
\end{abstract}

\pacs{71.10.Pm, 71.10.Hf, 73.21.Hb} 
\maketitle

Few systems are more universal than electron gases. Their study started
long-time ago and the compilation of knowledge that we have now at hand is
very wide, with impressive quantitative and qualitative
results~\cite{giuliani}. Phase diagrams for the electron gas in two and
three dimensions appear now quite well understood thanks to progressively
more accurate many-body calculations using mainly quantum Monte Carlo
methods~\cite{ceperley_rev}. However, the theoretical knowledge of the
electron gas in the one-dimensional (1D) geometry is more scarce and a full
determination of the density-temperature phase diagram is still lacking.
The present work is intended as a contribution towards filling this gap by
means of a microscopic approach based on the path integral Monte Carlo
(PIMC) method.

The quasiparticle concept introduced by Landau in his Fermi liquid theory
is able to account for the excitations of the electron gas in two and three
dimensions. This is not the case in one dimension where the enhancement of
correlations makes all excitations, even at low energy, to be collective.
The appropriate theoretical framework is an effective low-energy
Tomonaga-Luttinger (TL) theory~\cite{tomonaga,luttinger,haldane}, properly
modified by Schulz~\cite{schulz} to account for the long-range nature of
the Coulomb interaction. Probably, the most noticeable prediction of the TL
theory is the  separation between spin and charge degrees of freedom, whose
excitations are predicted to travel at different velocities. At the same
time, a Coulomb wire is fundamentally different from other TL systems in
that at low densities it forms a Wigner crystal, as manifested by the
emergence of quasi-Bragg peaks~\cite{schulz}. Also the strongly repulsive
nature of interactions might lead to a formation of a Coulomb
Tonks-Girardeau gas\cite{CTG}. In spite of the experimental difficulties in
getting real 1D environments, strong evidences of having reached the TL
liquid and the 1D Wigner crystal have been reported in the last
years~\cite{pouget, goni,auslaender,kim,jompol,laroche,deshpande}.
Therefore, the continued theoretical interest on this 1D system is
completely justified and can help to understand future experimental
findings.

The ground-state properties of the 1D Coulomb gas have been studied in the
past using several methods, the most accurate results being obtained using
the diffusion Monte Carlo (DMC) method~\cite{casula1,casula2,lee}. One of
the main goals of these calculations was the estimation of the interaction
energy of the gas with as higher precision as possible to generate accurate
density functionals to be used within density functional theory of
quasi-one-dimensional systems. All these calculations have been carried out
assuming a quasi-1D geometry imposed by a tight transverse confinement,
normally of harmonic type. In the latter case, one assumes that electrons
occupy the ground-state of the transverse harmonic potential and so in the
resulting effective Coulomb interaction the divergence at $x=0$ is
eliminated. Proceeding in this way, the effective one-dimensional
interatomic potential can be Fourier transformed. However, a recent DMC
calculation~\cite{astra} has shown that the use of the bare Coulomb
interaction is not a problem for the estimation of the energy and
structural properties because the wave function becomes zero when $|x|
\rightarrow 0$. More importantly, the presence of a node at $x = 0$ makes
Girardeau's mapping applicable~\cite{girardeau} which means that the
many-particle bosonic wave function is the absolute value of the fermionic
one, with the same Hamiltonian. In other words, the non-integrable
divergence of the interaction at small distances acts effectively as a
Pauli principle for bosons. From the computational point of view, this is
highly relevant because knowing the exact position of the nodes allows us
to perform an exact simulation without the usual upper-bound restriction
imposed by the fixed-node approximation when the nodal surfaces are
unknown.

We consider a system composed of $N$ particles with charge $e$ and mass $m$
in a 1D box of length $L$ with periodic boundary conditions, that interact
by means of a pure Coulomb potential. We work in atomic units, the Bohr
radius $a_0 = \hbar^2 / (m e^2)$ for the length and the Hartree $ \text{Ha}
= e^2 / a_0$ for the energy. In these reduced units, the Hamiltonian is
given by
\begin{equation}
\label{Eq:hamat}
H = - \frac{1}{2}\sum_{i=1}^{N} \frac{\partial^2}{\partial x_i^2} +
\sum_{i<j}^{N} \frac{1}{|x_i - x_j|} \ .
\end{equation}
At finite temperature $T$, the knowledge of the partition function $Z =
\mathrm{Tr} \, e^{-\beta H}$ ($\beta=1/T$) gives access to a microscopic
description of the properties of the system. The partition function
satisfies a convolution property which allows for its estimation via a path
integral Monte Carlo (PIMC) scheme,
\begin{equation}
Z = \int d{\bf R}_1 \ldots d{\bf R}_M \prod_{\alpha=1}^{M} \rho ({\bf
R}_\alpha,{\bf R}_{\alpha +1}) \ ,
\label{Eq:partition}
\end{equation}
where $\rho ({\bf R}_\alpha,{\bf R}_{\alpha +1})$ stands for an
approximated density matrix at higher temperature $M T$. In the present
work, we use a fourth-order approximation for $\rho$~\cite{Chin} that has
already proved its efficiency in the study of other systems such as liquid
$^4\mathrm{He}$~\cite{HighOrder}. Here, the fourth-order dependence of the
energy on $1/M$ is recovered at larger $M$ values than in $^4\mathrm{He}$
due to the pathological behavior of the Coulomb potential for the lowest
approximation for the action (primitive approximation)~\cite{broughton}.
Nevertheless, the high-order PIMC method is able to explore the major part
of the density-temperature phase diagram with accuracy and without any bias
coming from the fixed-node constraint.

Our main goal is the calculation of the phase diagram of the 1D Coulomb
quantum wire. To this end, we mainly determine the energetic and structure
properties of this system. For the energy we use the virial estimator,
which relies on the invariance of the partition function under a scaling of
the coordinate variables, thus providing good results at large values of
$M$, where the thermodynamic estimator fails to provide converged
results~\cite{HighOrder}. The structure properties of the system are
obtained from the behavior of the static structure factor, $S(k)=N^{-1}
\langle \hat{\rho}_k \hat{\rho}_{-k}  \rangle$, with $\hat{\rho}_k
=\sum_{l=1}^N \exp(i k x_l)$ the density operator.

\begin{figure}
\begin{center}
\includegraphics[width=0.8\linewidth,angle=0]{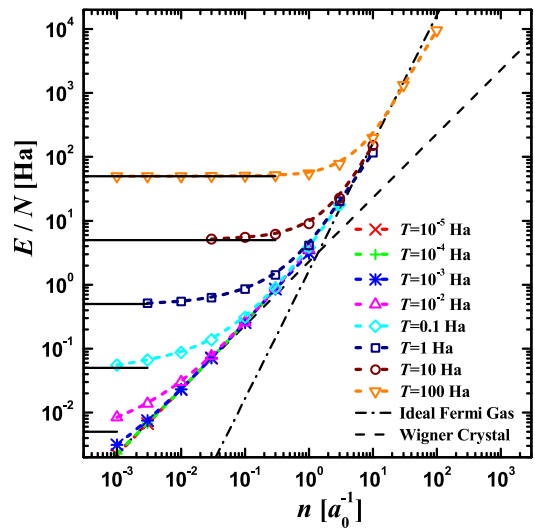}
\caption{(Color online) Energy per particle as a function of the density for $N = 10$.
Symbols, PIMC results at different temperatures; dashed line, energy of a Wigner crystal at $T = 0$; dash-dotted line, IFG energy at $T = 0$. Solid lines, classical limit $E/N = T/2$.
}
\label{Fig:energy diagram}
\end{center}
\end{figure}

The energies obtained at different temperatures and densities are shown in
Fig.~\ref{Fig:energy diagram}.  When both the temperature and density are
low, the potential energy dominates and the total energy can be estimated
by summing up all pair Coulomb potential energies for a set of particles at
the fixed positions of a Wigner crystal.  For a given number of particles
$N$, the leading term in the energy is linear with the density
$n$~\cite{dubin},
\begin{equation}
\frac{E_{\text W}}{N} = e^2 n \ln N    \ .
\label{wignerenergy}
\end{equation}
If one fixes the density and changes the number of particles,
Eq.~(\ref{wignerenergy}) predicts an energy per particle which diverges
logarithmically with $N$.  This is, in fact, a well known effect of the
long-range behavior of the Coulomb potential in strictly 1D
problems~\cite{astra}.  When the density increases, the kinetic energy
increases faster than the potential energy due to its quadratic dependence
with $n$.  Then, the system reaches a regime where the energy is well
approximated by the ground-state energy of an ideal Fermi gas,
\begin{equation}
\frac{E_{\text{ IFG}}}{N} = \frac{\hbar^2 k_F^2}{6 m}    \ ,
\label{fermienergy}
\end{equation}
with $k_F= \pi n$ being the 1D one-component Fermi momentum. Both limiting
behaviors, $E_{\text W}/N$ (\ref{wignerenergy}) and $E_{\text{ IFG}}/N$
(\ref{fermienergy}) are shown as straight lines which cross at $n \simeq 1$
in the log-log plot of Fig.~\ref{Fig:energy diagram}. The ground-state
energy obtained with the DMC method for $T=0$  is recovered in our PIMC
simulation when the temperature drops below some critical temperature,
which value depends on the density. Increasing the density in the ground
state, the system  evolves from a Wigner crystal to a zero-temperature
ideal Fermi gas~\cite{astra}. For a fixed finite temperature, $T\lesssim 1$
Ha,  the dilute regime of low density corresponds to a classical gas with
the energy per particle given by the classical value $E_{\text C}=T/2$
(solid horizontal lines) , the Wigner crystal is realized at larger
densities and, finally, the quantum wire behaves as an ideal Fermi gas for
$n \gtrsim 1$.   For temperatures $T \gtrsim 1$ Ha, the Wigner crystal
behavior is no more observed and the system evolves directly from a
classical gas to a Fermi one.

\begin{figure}
\begin{center}
\includegraphics[width=0.8\linewidth,angle=0]{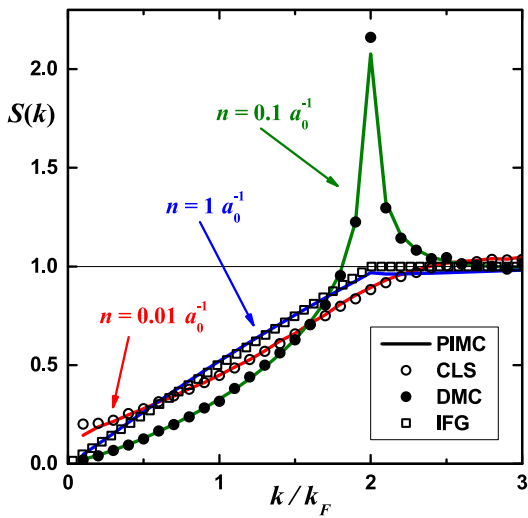}
\caption{(Color online) Static structure factor for $N=20$ and $T = 10^{-2}$ Ha for different densities. 
At the lowest density, we compare the PIMC result with a classical Monte
Carlo simulation (CLS). 
For densities $n \geq 0.1\,a_0^{-1}$, we make a comparison with the ground-state properties obtained by the DMC method  ($T=0$).
For densities $n \geq 1\,a_0^{-1}$, we compare the PIMC results with the Ideal Fermi Gas.
}
\label{Fig:TconstantTrans}
\end{center}
\end{figure}

In spite of the absence of real phase transitions in this 1D system one can
identify different regimes with well-known limiting cases. As we have shown
in Fig.~\ref{Fig:energy diagram}, the energy shows a rich variety of
behaviors as both the density and temperature are changed. However, it is
the study of the structural properties which provides us a deeper
understanding on the difference between regimes. To this end, we use the PIMC
method to calculate the density and temperature dependence of the
static structure factor $S(k)$.  Its behavior at a constant temperature
($T = 10^{-2}$ Ha)  and different densities is shown in
Fig.~\ref{Fig:TconstantTrans}. At the lowest density $n=0.01\,a_0^{-1}$, the quantum
PIMC results are nearly indistinguishable of the classical $S(k)$ obtained
by the classical Monte Carlo method (Boltzmann distribution) at the same
density and temperature. Increasing more the density, the static structure
factor shows clearly the emergence of a Bragg peak at $k/k_F=2$ signaling
the formation of a Wigner crystal in 1D~\cite{schulz}. At low temperatures,
the quantum degeneracy is reached and $S(k)$ agrees with that of a DMC
estimation at $T=0$ at the same density. Similarly to what happens at zero
temperature~\cite{astra}, increasing even more the density the system
evolves to an ideal Fermi gas.  In Fig.~\ref{Fig:TconstantTrans}, we also
compare the PIMC result for $S(k)$ at $n=1\,a_0^{-1}$ with the IFG $S(k)$  at the
same density and $T=0$: the agreement between both curves is excellent.

\begin{figure}
\begin{center}
\includegraphics[width=0.8\linewidth,angle=0]{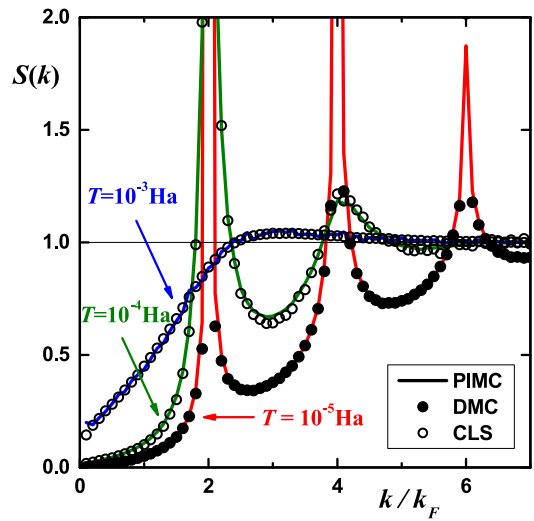}
\caption{(Color online) Static structure factor for $N=20$ and $n= 10^{-3}$ for different temperatures. 
At the lowest temperature we compare the PIMC result with the DMC result of $T=0$. 
At high temperature, we compare the PIMC result with a classical Monte
Carlo simulation (CLS).
}
\label{Fig:nconstantTrans}
\end{center}
\end{figure}

It is important to understand how the temperature affects the structural
properties,  when the density is fixed and the temperature is progressively
increased. Figure~\ref{Fig:nconstantTrans} reports PIMC results obtained at
low density, $n=10^{-3}\,a_0^{-1}$.  At low temperatures, one identifies the
characteristic Bragg peaks at $k/k_F= 2 l$ with integer $l$. At the lowest
considered temperature, $T=10^{-5}$ Ha, we observe a quantum crystal and
$S(k)$ is in nice agreement with the $T=0$ result obtained by the DMC
method. Increasing the temperature by a factor of ten, the presence of
Bragg peaks confirms the formation of  a Wigner crystal while its structure
is very different from the quantum one, observed at $T=0$. Importantly, we
find out that the correlations at  $T=10^{-4}$ Ha are the same as in a crystal
with electrons obeying Boltzmann statistics. 

Once in the classical regime, by increasing the temperature the crystal
\textit{melts} and becomes a gas.  In Fig.~\ref{Fig:nconstantTrans}, one
can observe that PIMC and classical simulations predict the same $S(k)$ in
a gas at temperature $T=10^{-3}$ Ha.  It becomes clear from
Figs.~\ref{Fig:TconstantTrans} and \ref{Fig:nconstantTrans} that the
transition between different regimes can be induced by changing the density
or the temperature.

\begin{figure}
\begin{center}
\includegraphics[width=0.8\linewidth,angle=0]{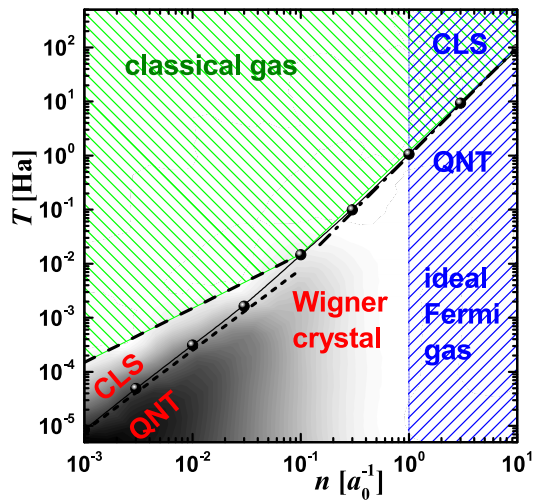}
\caption{(Color online) Temperature --  density \textit{phase} diagram.
Long-dashed line, gas-Wigner crystal crossover; 
dash-dotted line, locates the crossover between a quantum Fermi gas and a classical 
thermal gas ($T_F \propto n^2$); short-dashed line separates the 
classical (CLS) and quantum (QNT) regimes 
within the Wigner crystal, $T \propto n^{3/2}$;
symbols connected with a thin line (guide to an eye), position of the 
classical-quantum crossover estimated as $E_{\text{kin}} = T$.   
Ideal Fermi gas and Wigner crystal regimes for the considered number of particles 
are separated by $n \approx 1$. 
Within the Wigner phase, the ratio of the peak value in $S(k)$ for $N=20$ and $N=10$ 
is shown with a contour plot (white color, no difference; black, large difference).
}
\label{Fig:phase diagram}
\end{center}
\end{figure}

From the PIMC results for both energy and structure we establish the
temperature-density \textit{phase} diagram of the 1D Coulomb wire. The
phase diagram is reported in Fig.~\ref{Fig:phase diagram} and constitutes
the main result of our work.  We identify three different regimes:
classical Coulomb gas, Wigner crystal and ideal Fermi gas, where the last
two regimes show a crossover from quantum to classical behavior. The Wigner
crystal is identified by calculating the ratio of the peak's height of
$S(k)$ at $k/k_F=2$ for two values of the number of particles ($N=20$, 10).
When the height of the peak increases with $N$, the system behaves as a
Wigner crystal. In Fig.~\ref{Fig:phase diagram}, we use a contour plot to
show that ratio in a grey scale, where black color stands for large ratio
and white for ratio equal to one. In the $T$-$n$ plane, the Wigner crystal
phase shows a triangular shape, with the strongest signal localized in the
vertex of lowest density and temperature, delimited by transitions to a
Coulomb or an ideal Fermi gas.  This quantum Wigner crystal is well
described by the zero-temperature theory, as we have shown in Figs.
\ref{Fig:TconstantTrans} and \ref{Fig:nconstantTrans}. Increasing the
temperature, one can see how the quantum crystal transforms into a
classical Wigner lattice.  In both regimes, particles move around the
lattice points but these fluctuations are of quantum and thermal nature in
quantum and classic crystals, respectively.  Starting from a
high-temperature crystal and by decreasing the temperature we see that the
system becomes more ordered and the height of the peaks increases. Indeed,
at zero temperature the classical system would always form a perfect
crystal with no fluctuations. Instead, we see that the height of the peaks
stops growing when we decrease the temperature down to the
quantum-degeneracy regime. By lowering the temperature further the system
remains in the ground state. In fact, the classical crystal regime is
realized when the temperature is large compared to the height of the first
Brillouin zone ($E_{\text{BZ}}$),  $E_{\text{BZ}} \ll T$. That can be
estimated from the phonon spectrum  at the border of the Brillouin zone,
$E_{\text{BZ}} = E_{\text{ph}}(k_{\text{BZ}}) \approx \hbar c
|k_{\text{BZ}}|$, with $k_{\text{BZ}} = 2 k_F =2 \pi n $. The speed of
sound $c$ is related to the chemical potential through the compressibility
relation $mc^2 = n\partial \mu /\partial N$.  As $\mu$ in the Wigner
crystal is linear in $n$, one can locate the transition from the quantum to
the classical Wigner crystal as $T \sim n^{3/2}$ (short-dashed line
in Fig.~\ref{Fig:phase diagram}). When the temperature is  high enough,
thermal fluctuations become large compared to the potential energy of the
Coulomb crystal and thus the Wigner crystal melts to a classical Coulomb
gas. As the energy of the Wigner crystal is linear with the density (for a
fixed number of particles $N$) (\ref{wignerenergy}) this \textit{melting}
transition line follows approximately the law $T \sim n$ (dashed line
in Fig.~\ref{Fig:phase diagram}).

By changing the density while keeping the temperature fixed to a very low
value, the system evolves from a Wigner crystal towards an ideal Fermi gas.
This evolution is driven by the different dependence of the potential and
kinetic energies on the density.  The kinetic energy grows quadratically,
$E_{\text kin}/N \propto n^2$,  instead of the linear dependence of the
potential energy, $E_{\text W} \propto n \ln N$.  At $n \approx 1$, we
observe this transition both in energy and in the shape of the static
structure factor $S(k)$.  For temperatures smaller than $T\lesssim 10^{-2}$
Ha,  we observe two different transitions: at low densities an evolution
from a thermal classical gas to a Wigner crystal, and at $n \approx 1$ the
\textit{melting} of the crystal towards the Fermi gas.  When $T>10^{-2}$
Ha, the Wigner crystal is no more stable and the evolution with the density
is from a classical gas to an IFG for densities $n > 1$.  On the other
hand, the finite-size dependence is very weak as it can be appreciated from
the logarithmic dependence of the Wigner crystal energy on $N$,
Eq.~(\ref{wignerenergy}).  Still, it becomes important when the number of
electrons is large.  It is expected that the stability region  of the
Wigner crystal will increase with $N$ both in density~\cite{astra} and in
temperature. 

The transition from the zero-temperature ideal Fermi gas to a classical gas
is governed by a single parameter, namely the ratio of the temperature and
the Fermi temperature, $T/T_F = T / [\pi^2n^2/2m]$.  When this ratio is
much smaller than one, the system stays in the ground-state of a quantum
degenerate gas.   When this ratio is much larger than one, the energy
approaches that of a Boltzmann classical gas.   In between, the system
properties are that of a finite-temperature quantum ideal Fermi gas.   A
special feature of the one-dimensional world is that the stability of the
quantum degenerate regime is greatly increased.   Indeed, the stability
regime grows rapidly as the density is increased since $T/T_F \propto
n^2$.  This should be contrasted with $T/T_F \propto n$ in two dimensions
and even weaker $n^{2/3}$ dependence in three dimensions. 

Summarizing, we have carried out a complete PIMC study of the
density-temperature phase diagram of a 1D quantum Coulomb wire.  The
singularity of the Coulomb interaction at $x=0$ allows us to solve the sign
problem and makes it possible to carry out an exact calculation of the
electron gas problem since we know a priori the exact position of the Fermi
nodes.  This is clearly a special feature of the 1D environment which
cannot be translated to higher dimensions.  There, in 2D and 3D, one can
only access to approximate solutions to the many-body problem which worsen
when the the temperature is not zero.  Focusing our analysis on energetic
and structural properties we have been able to characterize the different
regimes of the electron wire.  In spite of the lack of real phase
transitions due to the strictly 1D character of the system, we have been
able to define different physical regimes, including the Wigner crystal 
(classical and quantum), the classical Coulomb gas, and the universal ideal
Fermi gas.  Two relevant features make this phase diagram specially
interesting: the large stability domain of the ideal Fermi gas and the
double crossing gas-crystal-gas with increasing density within a quite wide
temperature window.  Our results are relevant to current and future
experiments with electrons in a quantum wire and to Coulomb ions  in
one-dimensional confinement.

We acknowledge partial financial support from the MICINN (Spain) Grant
No.~FIS2014-56257-C2-1-P.  The Barcelona Supercomputing Center (The Spanish
National Supercomputing Center -- Centro Nacional de Supercomputaci\'on) is
acknowledged for the provided computational facilities.

\end{document}